\documentstyle[aps,prb,amssymb,twocolumn,epsf]{revtex}

\begin{document}

\title{Effective Ginzburg-Landau model for superfluid $^4$He}

\author{Torsten Fliessbach\cite{email}}

\address{Fachbereich Physik, University of Siegen, 
 57068 Siegen, Germany}


\maketitle

\begin{abstract}

We propose an effective Ginzburg-Landau energy functional for
superfluid helium that is motivated by the ideal Bose gas model.
The parameters of this phenomenological model are adjusted to known
properties of the bulk system. The motivation and the field of
applications of this model are quite similar to that of the $\Psi$
theory by Ginzburg and Sobyanin. The correlation length, the
depression of the transition temperature in helium films and the
density profile at a boundary are calculated. 
\end{abstract}


\section{Introduction}
\label{s1}

In the Ginzburg-Landau (GL) model for liquid $^4$He the superfluid
density $\varrho_{\rm s}$ is identified with the expectation value of
the square of the complex order parameter field $\Psi({\bf r})$,
\begin{equation}
\label{e1}
\varrho_{\rm s} = m\left\langle |\Psi|^2 \right\rangle \ .
\end{equation}
Here $m$ is the mass of a helium atom. In the vicinity of the
$\lambda$-transition, the energy functional is expanded into powers
of the field $\Psi$ and its derivatives and of the relative temperature
$t=T/T_{\lambda} - 1$.  In its simplest form the GL energy functional
takes into account the terms with $|\nabla \Psi|^2$, $t\cdot |\Psi|^2$
and $|\Psi|^4$.

The GL model for superfluid helium has been introduced in 1958 by
Ginzburg and Pitaevski\u{\i}\cite{gp58}. It leads to critical
properties (like $\varrho_{\rm s} \propto |t|$) that do not agree with
the experiment. This failure is due to the fact that in this model the
fluctuations diverge (relative to the mean value) when the transition
point is approached.

The theoretical remedy for this break-down of the GL model is
well-known: The GL ansatz may be considered as a sensible approximation
for finite volumes of the sample only. Using scale transformations
to larger and larger volumes renormalizes the coefficients in the
energy functional.  The renormalization group theory (RGT) leads to
nontrivial critical exponents that are no longer in variance with
experiment. RGT is considered as the best present-day theory which
allows in principle the calculation of all relevant quantities.

For practical reasons it is, however, useful to have a phenomenological
GL like model that is in accordance with the bulk properties and in
particular with the experimental critical exponents of superfluid
helium. Such an effective GL model allows to address a large variety of
problems in a simple and straight-forward way.

An effective GL model of this kind has been put forward under the name
``$\Psi$ theory'' by Ginzburg and Sobyanin\cite{gs76,gs82,gs88}. The
references\onlinecite{gs76,gs82,gs88} are comprehensive papers
addressing the foundation and applications of such an approach. For the
historical development of the field we refer to the corresponding
discussion and the bibliography in these papers.

The $\Psi$ theory by Ginzburg and Sobyanin (abbreviated by GS in the
following) is based on Eq.\ (\ref{e1}), too. The GS energy functional
contains the terms $|\nabla \Psi|^2$, $|t|^{4/3}\cdot |\Psi|^2$,
$|t|^{2/3}\cdot |\Psi|^4$ and $|\Psi|^6$.  The temperature dependences
of the coefficients are chosen by hand such that they yield
approximately the right critical exponents (like $\varrho_{\rm s}
\propto |t|^{2/3}$, for example). In so far as the actual critical
exponent of $\varrho_{\rm s}$ deviates from $\nu=1/3$ this model
is an approximation. For the sensibility and usefulness of such an
effective GL model and its relation to RGT we refer to the thorough
discussion by GS\cite{gs76,gs82,gs88}.

It is generally accepted that there is an intimate
connection\cite{fe53,lo54,pu74} between the Bose-Einstein condensation
of the ideal Bose gas (IBG) and the $\lambda$ transition in liquid
helium.  In Ref.\onlinecite{fl99} we have argued that there are good
reasons to uphold the IBG critical exponent $\beta=1/2$ of the
condensate density $\varrho_0 = m\,\langle |\Psi|^2 \rangle \sim |t|$.
This implies $\varrho_0 <\varrho_{\rm s}$ just below the lambda point.
Consequently, we have proposed that the superfluid density is composed
by the condensate density and a coherently comoving density
$\varrho_{\rm coh}$ (corresponding to lowlying noncondensed states):
\begin{equation}
\label{e2}
\varrho_{\rm s} = \varrho_0 + \varrho_{\rm coh}
\ ,\quad\mbox{where}\quad
\varrho_0 = m \left\langle |\Psi|^2\right\rangle
\ .
\end{equation}
The arguments in favor of this composition of $\varrho_{\rm s}$ and the
resulting expression for $\varrho_{\rm s}$ are reviewed in Sec.\
\ref{s2.1}. The expression for $\varrho_{\rm s}$ provides excellent
fits\cite{fl99} to the experimental temperature dependence of the
superfluid density.

Based on the starting point (\ref{e2}) instead of Eq.\ (\ref{e1}) we
construct in this paper an alternative effective GL model. Because of
$\varrho_0 = m\,\langle |\Psi|^2\rangle\sim |t|$ we may use the
standard form for the Landau part of the energy functional (i.\,e.\/
the terms $t\cdot |\Psi|^2$ and $|\Psi|^4$). The kinetic energy
contribution (the term with $|\nabla \Psi|^2$) gets, however, an
additional factor $\varrho_{\rm s}/\varrho_0$ due to the coherent
comotion. Effectively, this factor $\varrho_{\rm s}/\varrho_0 \sim
|t|^{-1/3}$ damps the critical fluctuation such that our GL ansatz
becomes scaling invariant.

From a pragmatic point of view our effective GL model (with the terms
$|t|^{-1/3}\cdot|\nabla \Psi|^2$, $t\cdot |\Psi|^2$ and $|\Psi|^4$) is
quite comparable to the $\Psi$ theory by GS (with the terms $|\nabla
\Psi|^2$, $|t|^{4/3}\cdot |\Psi|^2$, $|t|^{2/3}\cdot |\Psi|^4$ and
$|\Psi|^6$). The main difference is that in our model the Landau part
is of its standard form (in accordance with the IBG), and that specific
reasons are given for the occurrence of a fractional power of $|t|$.

This paper is organized as follows. Section \ref{s2} presents the
reasons for the composition (\ref{e2}) of the superfluid density,
the actual expression for $\varrho_{\rm s}$, and the parameters
obtained by a fit to the experimental temperature dependence of
$\varrho_{\rm s}$. Section \ref{s3} displays the energy functional
of our effective GL model, and discusses its basic features.  Section
\ref{s4} investigates the temperature dependence of the correlation
length and the size of the critical fluctuations. The depression of the
transition temperature in helium films is calculated in Sec.\ \ref{s5}.
Section \ref{s6} derives the spatial variation of the superfluid and
the condensate density at a boundary.

\section{Composition of the superfluid density}
\label{s2}

\subsection{Motivation}
\label{s2.1}

The comparison between liquid $^3$He and $^4$He on one side and the
ideal Fermi gas and the ideal Bose model (IBG) on the other side
strongly suggests that there is an intimate relation between the
Bose-Einstein condensation (BEC) and the lambda transition in liquid
$^4$He. It is, moreover, inviting to identify $\Psi$ in Eq.\ (\ref{e1})
with the IBG condensate wave function. This identification
explains\cite{lo54,pu74} a number of experimental findings of which the
most important one \cite{no90} is that a superfluid current is
irrotational. The IBG {\em and}\/ this identification are, however, in
conflict with the experimental critical exponent $\nu\approx 1/3$ of
$\varrho_{\rm s}\sim |t|^{2\nu}$ (as compared to the IBG with
$\beta=1/2 $ for $\langle|\Psi|^2\rangle \sim |t|^{2\beta}$).

The standard solution of this conflict appears to be the
renormalization group theory (RGT). The RGT yields indeed an
explanation of the experimental value of $\nu\approx 1/3$. It does,
however, not resolve the conflict with the IBG value $\beta=1/2$: A
renormalization is appropriate for the Landau value $\beta_{\rm L}
=1/2$ but not for the IBG value $\beta=1/2$. The reason is that the IBG
value is obtained by the exact evaluation of a partition sum. This
exact evaluation implies a summation over arbitrarily small momenta
(or, correspondingly, arbitrarily large distances). Therefore, the
reasoning behind the renormalization procedure (analytic
Ginzburg-Landau ansatz for {\em finite}\/ regions or blocks, and
subsequent transformation to larger and larger blocks) cannot be
applied to the IBG energy functional.

Within the framework of the ideal Bose gas model we have
proposed\cite{fl99} to resolve the conflict $\beta=1/2$ and $\nu\approx
1/3$ by the assumption that {\em noncondensed particles move coherently
with the condensate}\/.  This means that we no longer identify the
condensate with the superfluid fraction; the condensate is only part of
the superfluid phase. The coherent comotion is constructed in the
following way: The condensate wave function may be written as
\begin{equation}
\label{e3} 
\Psi({\bf r}) = \varphi_0({\bf r}) \exp [ \,{\rm i}\hspace*{0.2ex}\Phi
({\bf r}) ]\ ,
\end{equation}
where $\varphi_0({\bf r})$ and $\Phi({\bf r})$ are real functions. The
velocity of a potential superfluid flow is given by ${\bf v}_{\rm
s} =(\hbar/m)\nabla \Phi$. The single particle function
$\varphi_{\bf k}({\bf r})$ of the noncondensed states can be chosen as
real functions. If now the lowlying states (up to a coherence
limit $k_{\rm coh}$ that is to be determined later) adopt the same
phase factor,
\begin{equation}
\label{e4} 
\varphi_{\bf k}({\bf r}) \;\to\; \varphi_{\bf k}({\bf r})
 \exp [\,{\rm i}\hspace*{0.2ex} \Phi({\bf r})] \ ,
\qquad (k\le k_{\rm coh})
\end{equation}
then the corresponding particles contribute to the superfluid current
density. Consequently, the superfluid density is made up by the
condensate density $\varrho_0$ and the density $\varrho_{\rm coh}$ of
the coherently comoving particles:
\begin{equation}
\label{e5}
\frac{\varrho_s}{\varrho} = \frac{1}{N}\,\left( \langle n_0\rangle +
{\sum_{k\le k_{\rm coh}}}^{\!\!\!\prime} \langle n_k\rangle
\right) = \frac{\varrho_0+\varrho_{\rm coh}}{\varrho} \ .
\end{equation}
Here $\langle n_k\rangle$ are average occupation numbers of the
(slightly modified) IBG, and the prime at the summation symbol excludes
the condensed particles from this sum.

We emphasize that the construction (\ref{e3}) to (\ref{e5}) preserves
the basic properties of the superfluid flow (like its irrotational
behavior).  In section \ref{s3.3} we discuss the physical meaning of
$k_{\rm coh}$ and present an argument why the lowest single particle
levels might adopt the same phase factor as the condensate.

\subsection{Functional form of the superfluid density}
\label{s2.2}
We determine the asymptotic form of the expression (\ref{e5}). The
IBG expectation values $\langle n_k\rangle$ are given by
\begin{equation}
\label{e6}
\langle n_k\rangle =
\frac{1}{\exp \hspace*{0.2ex} [ (\hspace*{0.2ex}
\varepsilon_k-\mu )/k_{\rm B}T] -1} 
= 
\frac{1}{\exp (\kappa^2+\tau^2)-1}
\ .
\end{equation}
Here $\mu$ is the chemical potential, $\varepsilon_k=\hbar^2 k^2/2m$
are the single-particle energies, and $k_{\rm B}$ is Boltzmann's
constant.  In the last expression we introduced the dimensionless momentum
\begin{equation}
\label{e7}
\kappa = \frac{\lambda \,|{\bf k}|}{\sqrt{4\pi }}\ ,
\quad\mbox{where }\,\,
\lambda =\frac{2\pi\hbar}
{\sqrt{2\pi\hspace*{0.2ex} m \hspace*{0.2ex}k_{\rm B}T}}
\ .
\end{equation}
The chemical potential $\mu$ has been expressed by the dimensionless
quantity $\tau^2=-\mu /k_{\rm B}T$  in Eq.\ (\ref{e6}). We expand this
$\tau$ into powers of the relative temperature $t = T/T_{\lambda}-1$:
\begin{equation}
\label{e8}
\tau (t) =\sqrt{\frac{-\mu}{k_{\rm B}T}} = 
\left\{ \begin{array}{lll}
a\, t+b \,t^2 +\ldots &\quad & (t>0) \\[1.5mm] 
a'\,|t| + b'\,t^2+\ldots
&&(t<0)
\end{array} \right.
\ .
\end{equation}
For $t>0$ the particle number condition $N=\sum_{\bf k}\langle n_{\bf
k}\rangle$ determines the temperature dependence of $\tau (t)$ and in
particular the coefficients $a$, $b$,\ldots{}, for example $a=
3\,\zeta(3/2)/(4\hspace*{0.2ex} \pi^{1/2})$. (Here $\zeta(x)
=\sum_1^\infty n^{-x}$ denotes Riemann's zeta function). For $t<0$ the
IBG yields $\tau=0$. As a modification of the IBG we admit nonvanishing
coefficients $a'$, $b'$,\ldots{}. By $a'\ne 0$ we introduce a
phenomenological gap between the condensate level and the noncondensed
levels. This modification does not affect the BEC as the most
important feature of the IBG. It is necessary in order to reproduce the
experimental temperature dependence of the superfluid density by
Eq.\ (\ref{e5}). In view of the successful roton picture it is not
surprising that a gap is required for a quantitative description.

For $t<0$ the particle number condition $N = \langle n_0 \rangle+
\sum'_{\bf k}\langle n_{\bf k}\rangle$ leads to
\begin{equation}
\label{e9}
\frac{\varrho_0}{\varrho} = \frac{\langle n_0\rangle}{N} = 
f \hspace*{0.2ex} |t| + {\cal O}(t^2)\ ,
\end{equation}
where
\begin{equation}
\label{e10}
f = \frac{3}{2} + \frac{2\sqrt{\pi }\,a'}{\zeta (3/2)}
\ .
\end{equation}

An evaluation of the coherently comoving density in (\ref{e5}) yields
\begin{equation}
\label{e11}
\frac{\varrho_{\rm coh}}{\varrho} =\frac{1}{N}\,
{\sum_{k \le k_{\rm coh}}}^{\!\!\!\prime} \langle n_k\rangle
 = \frac{4\,\kappa_{\rm coh}}{\sqrt{\pi}\, \zeta(3/2)} 
+ \mbox{higher order} 
\ .
\end{equation}
The leading experimental behavior is approximately
\begin{equation}
\label{e12}
\frac{\varrho_{\rm s}}{\varrho} \approx \frac{\varrho_{\rm coh }}{\varrho}
\approx a_0\hspace*{0.2ex} |t|^{2/3} \ .
\end{equation}
Deviations from the exponent 2/3 are neglected in our phenomenological
model. Equation (\ref{e12})  corresponds to
\begin{equation}
\label{e13}
\kappa_{\rm coh}=\kappa_0 \hspace*{0.2ex}|t|^{2/3}\ ,
\end{equation}
where
\begin{equation}
\label{e14}
a_0 = \frac{4\,\kappa_0}{\sqrt{\pi}\,\zeta(3/2)} \ .
\end{equation}

\subsection{Results}
\label{s2.3}

An excellent agreement with the experimental temperature dependence of
the superfluid density is obtained\cite{fl99} if higher order terms in
Eqs.\ (\ref{e9}) and (\ref{e11}) are included. The resulting fit
formula contains four parameters as do comparable standard fits.

\begin{figure}[h]
\begin{center}
\epsfxsize=8cm
\epsfbox{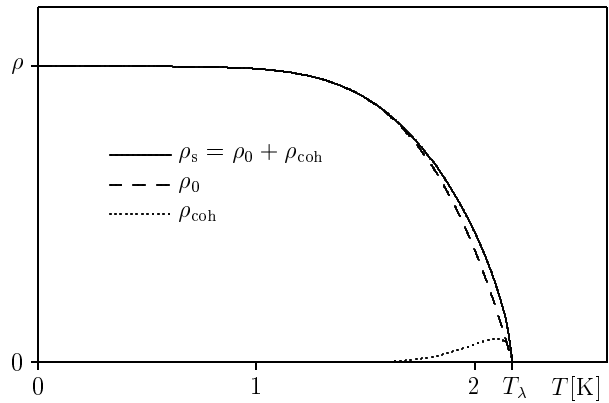}
\end{center}
\caption[]{\label{fig1}Decomposition of the superfluid density
$\rho_{\rm s}$ into the model condensate density $\rho_0$ and the
coherently comoving density $\rho_{\rm coh}$ as a function of the
temperature $T$. For $T\to T_\lambda$ the comoving density $\rho_{\rm
coh}\propto |t|^{2/3}$ is the dominant contribution to the superfluid
density.}

\end{figure}

This fit defines the temperature dependence of the contributions
$\varrho_{\rm coh}$ and $\varrho_0$ to the superfluid density. The
result is shown in Fig.\ \ref{fig1} that has been taken from
Ref.\onlinecite{fl99}. For $|t|<0.01$ one finds $\varrho_{\rm
s}\approx \varrho_{\rm coh}$ (because of $\varrho_{\rm coh}\sim
|t|^{2/3}$ and $\varrho_0\sim |t|$). Due to the gap between the
condensed and the noncondensed particles, the contribution
$\varrho_{\rm coh}$ is substantial only in a narrow region ($|t|<0.2$).
For $|t|>0.2$ one obtains $\varrho_{\rm s}\approx
\varrho_0$.

The model condensate density $\varrho_0$ may be multiplied by a
depletion factor (roughly $0.1$) in order to become comparable to the
measurable condensate density. The temperature dependence of the
resulting condensate density compares well with experimental results by
Snow et al.\cite{sn92}.

In this paper we need only the coefficients of the leading terms in
Eqs.\ (\ref{e9}) and (\ref{e12}). The fit of the theoretical expression
to the experimental data yields the following parameter
values\cite{fl99}
\begin{equation}
\label{e15}
a'= 3.03 \ ,\quad \kappa_0 = 2.70
\ .
\end{equation}
According to Eqs.\ (\ref{e10}) and (\ref{e14}) these parameters are
equivalent to
\begin{equation}
\label{e16}
f = 5.60 \ ,\quad a_0 = 2.33 \ .
\end{equation}
These two parameters are used as an input for our energy functional.

\section{Effective Ginzburg-Landau functional}
\label{s3}

\subsection{Definition}
\label{s3.1}

The equilibrium value of Gibbs free energy $G$ depends on the
temperature $T$, the pressure $P$ and the total particle number $N$. In
a GL like model, the thermodynamic potential $G$ depends, moreover, on
the order parameter field $\Psi$. The minimum of $G$ with respect to 
$\Psi$ determines its expectation value. The equilibrium value of the
potential is then $G(T,P,N) = \langle G(\Psi)\rangle$.

The number of particles is conserved in the considered applications.
The pressure dependence will not be discussed explicitly; it enters via
$T_{\lambda}(P)$ and via the pressure dependence of the parameters
(\ref{e16}) that may be determined by the fit to the experimental
superfluid density $\varrho_{\rm s}(T,P)$. The following considerations
concentrate on the $t$ and $\Psi$ dependences.

Equations (\ref{e2}) and (\ref{e9}) yield the bulk equilibrium value 
\begin{equation}
\label{e17}
v \left\langle |\Psi(\mbox{bulk})|^2\right\rangle = 
\frac{\langle n_0\rangle}{N} = f\hspace*{0.2ex} |t| +\ldots
\; .
\end{equation}
Here $v=V/N$ is the volume per particle. The order parameter field
$\Psi({\bf r})$ will, in general, depend on the coordinates. The
thermodynamic potential is, therefore, written as an energy functional.

We split the thermodynamic potential into several parts:
\begin{equation}
\label{e18}
G = G_0 + G_{\rm GL} = G_0 + G_{\rm fluct} + G_{\rm L} \ .
\end{equation}
The contribution that is not directly related to the order parameter
field is denoted by $G_0$. Following GS\cite{gs76,gs82} we assume that
the approximately logarithmic singularity of the specific heat is due
to this term. Our investigation is restricted to the Ginzburg-Landau
(GL) contribution $G_{\rm GL}$ that is due to the order parameter
field. This contribution may be further split into the Landau part
$G_{\rm L}$ and a part $G_{\rm fluct}$ that contains derivatives of the
field $\Psi$.

Near the transition point the energy functional is expanded into powers
of the order parameter field:
\begin{equation}
\label{e19}
G_{\rm GL} = \int\! d^3r \, \left( A\, |\nabla\Psi|^2 
+ B\, |\Psi|^2 + C\,|\Psi|^4 + \ldots \,\right) \ .
\end{equation}
First we consider the terms without derivatives, i.\,e.\/ the Landau part
$G_{\rm L}$. The most simple standard Landau ansatz ($B=b\,t$, $C =
\mbox{const.}$ and no higher order terms) leads to the IBG like behavior
(\ref{e17}). The coefficient in Eq.\ (\ref{e17}) is reproduced if we
set $G_{\rm L}\propto t\,|\Psi|^2/f + v\hspace*{0.2ex}|\Psi|^4/(2f^2)$.
For the determination of the prefactor we follow GS\cite{gs76,gs82} who
argue that $G_{\rm L}$ should account for the jump of the specific heat
at the lambda point. This leads to
\begin{equation}
\label{e20}
\frac{G_{\rm L}}{k_{\rm B} T_{\lambda}} =  
\int\! d^3r \,\frac{\Delta c_P}{k_{\rm B}}\, \left(
t\,\frac{|\Psi|^2}{f} + \frac{v \hspace*{0.2ex} |\Psi|^4}{2f^2}\right)
\ .
\end{equation}
where $\Delta c_P$ is the jump of the specific heat per particle.
The bulk expectation value of the r.\,h.\,s.\/ for $t<0$ is $- V (\Delta
c_P/k_{\rm B}) \hspace*{0.2ex} |t|^2 /(2v)$ yielding $\langle
G_{\rm L}/N\rangle = - \Delta c_{\rm P}\hspace*{0.2ex} T_\lambda
\hspace*{0.2ex} |t|^2/2$. For $t>0$ we have $\langle G_{\rm L}/N\rangle
= 0$. The second temperature derivative of these expectation values
reproduces the jump of the specific heat.

The contribution $G_{\rm fluct}$ may be written as
\begin{equation}
\label{e21}
\frac{G_{\rm fluct}}{k_{\rm B} T_{\lambda}}
=  \frac{1}{k_{\rm B}T_{\lambda}} 
\int\! d^3r \,\frac{\hbar^2}{2 \hspace*{0.2ex} m}\, R\, |\nabla\Psi|^2
\ .
\end{equation}
In a standard GL ansatz the factor $R$ equals $1$. According to
Eqs.\ (\ref{e3})\,--\,(\ref{e5}) the density $\varrho_{\rm coh}$ moves
coherently with the condensate density $\varrho_0$.  The term
$|\nabla\Psi|^2$ corresponds to the product of $\varrho_0$ times a
squared velocity. This has to be replaced by $\varrho_{\rm s}$ times
the squared velocity. This implies
\begin{equation}
\label{e22}
 R(t) = \frac{\varrho_{\rm s} (\mbox{bulk}) }{\varrho_0(\mbox{bulk})}
\simeq \frac{a_0\hspace*{0.2ex}|t|^{2/3}}{f \hspace*{0.2ex} |t|}\,.
\end{equation}
For the last expression we used Eqs.\ (\ref{e9}) and (\ref{e12}). By
$R\sim |t|^{-1/3}$ we admit a fractional power of the relative
temperature in the coefficients (similarly as in the $\Psi$ theory by
GS).

From Eqs.\ (\ref{e20}), (\ref{e21}) and (\ref{e22}) we obtain the
central expression for the energy functional:
\begin{equation}
\label{e23}
\frac{G_{\rm GL}}{k_{\rm B} T_{\lambda}} = \int\! d^3r 
\left[ \frac{a_0 \hspace*{0.2ex} \lambda_{\rm c}^{\,2}\hspace*{0.2ex}
 |\nabla\Psi|^2}  {4\pi f \hspace*{0.2ex} |t|^{1/3}}
 +  \frac{\Delta c_P}{k_{\rm B}}
\left( \frac{t\,|\Psi|^2}{f} + \frac{|\Psi|^4}{2f^2}\right) \right] .
\end{equation}
Here $\lambda_{\rm c}$ denotes the thermal wavelength at the critical
temperature:
\begin{equation}
\label{e24}
\frac{\lambda_{\rm c}}{\sqrt{4\pi}}
 =\frac{\hbar}{\sqrt{2 \hspace*{0.2ex} m
\hspace*{0.2ex} k_{\rm B} T_{\lambda}}}
\approx 1.67\,\mbox{\AA}
\ .
\end{equation}
For the numerical value we inserted the mass $m=m(\mbox{He})$ of a
helium atom and $T_{\lambda} \approx 2.17\,{\rm K}$.  The other
coefficients in the energy functional are given by
\begin{equation}
\label{e25}
a_0 = 2.33 \ ,\quad
f   = 5.60 \ ,\quad
\frac{\Delta c_P}{k_{\rm B}} = 2.77 \ .
\end{equation}
The first two parameters are determined by the fit of Eq.\ (\ref{e5})
to the experimental superfluid density (Sec.\ \ref{s2.3}).  The value
for $\Delta c_P$ is taken from Table V (third line) of
Ref.\onlinecite{ah71}. To some extent, the parameter values depend the
other terms in the corresponding fit formula used.

We use preferably dimensionless quantities: Both sides of
Eq.\ (\ref{e23}) are dimensionless. The integral $\int\!d^3r\,|\Psi|^2$
is dimensionless and of order $N$. The product $\lambda _c\nabla$ is
dimensionless, too. The occurring numerical factors ($a_0$, $f$ and
$\Delta c_P/k_{\rm B}$) are of order $1$.

\subsection{Discussion}
\label{s3.2}

\subsubsection{Comparison to other approaches}
\label{s3.2.1}

We consider the functional
\begin{equation}
\label{e26}
G_{\rm GL} =  \int\! d^3r \, \left( A\, |\nabla\Psi|^2 
+ B\, |\Psi|^2 + C\,|\Psi|^4 + D\,|\Psi|^6  \right) \ .
\end{equation}
and compare the coefficients of our model with that of a standard
Ginzburg-Landau ansatz (GL) and of the $\Psi$ theory (GS):
\begin{equation}
\label{e27}
\begin{tabular}{c|cccc}
& $A$ & $B$ & $C$ & $~D$ \\ \hline 
GL\cite{gp58} & 1 & $t$ & 1 & ~0 \\
GS\cite{gs76,gs82,gs88} & 1 & ~$|t|^{4/3}$~ & ~$|t|^{2/3}$~ & ~1 \\
~this work~ & ~$|t|^{-1/3}$~ & $t$ & 1 & ~0 
\end{tabular}
\end{equation}
The entrance 1 stands for a nonvanishing constant. For GS the
$|\Psi|^6$ term must be included since it scales with $t^2$ (as do the
terms $|t|^{4/3}|\Psi|^2$ and $|t|^{2/3}|\Psi|^4$). Because of this
term the GS functional contains an additional adjustable parameter. In
the standard GL as well as in our approach the $|\Psi|^6$ term is a
higher order term.

Due to the connection to the IBG, our Landau part (coefficients $B$,
$C$ and $D$) coincides with the standard GL form. The occurrence of the
$|t|^{-1/3}$ dependence in the fluctuation term (coefficient $A$) is
due to the coherent comotion of noncondensed particle with the
condensate.

As far as the results are concerned, our approach is more similar to
GS than to GL. In particular the correlation length scales with
$|t|^{-2/3}$ as in the $\Psi$ theory, and not with $|t|^{-1/2}$ as for
GL.

\subsubsection{Choice of coefficients}
\label{s3.2.2}

The prefactor of the Landau part (\ref{e20}) has been chosen such that
the expectation value $\langle G_{\rm L}\rangle$ reproduces the jump
$\Delta c_P$ of specific heat at the lambda point. This choice may be
criticized for two reasons: (i) The fluctuation part $\langle G_{\rm
fluct}\rangle$ leads to a $t^2$ term for $t<0$, and will contribute to
the jump, too. (ii) There might be a contribution to the jump from the
part $\langle G_0\rangle$ in Eq.\ (\ref{e18}). By assuming that the
jump is solely due to the Landau part $\langle G_{\rm L}\rangle$ we
follow for simplicity GS. The alternative is to replace $\Delta
c_P/k_{\rm B}$ by a unknown parameter. Adjusting this parameter later
(for example to the correlation length amplitude) would yield a value
rather close to $\Delta c_P/k_{\rm B}$. This might be considered as a
pragmatic reasoning of the present choice.

The factor $R =\varrho_{\rm s}/\varrho_0$ introduced in Eq.\
(\ref{e21}) has been motivated by the coherent comotion (Eqs.\
(\ref{e3}) to (\ref{e5})) of noncondensed particles. This comotion
refers to a potential superfluid flow with ${\bf v}_{\rm s} = (\hbar/m)
\hspace*{0.2ex} \nabla \Phi({\bf r})$. The motivation for the
$R$-factor has thus been based on a rather specific coordinate
dependence of $\Psi({\bf r})$ for $t<0$. It is, however, the spirit of
a GL like approach to consider only one functional for the complex
$\Psi({\bf r})$ field, i.\,e.\ to use the same coefficients for all
occurring coordinate dependences, and for $t<0$ and $t>0$.
Alternatively, one could introduce different coefficients for the
coordinate variation of the amplitude  and the phase of the $\Psi({\bf
r})$ field; this would result in a more complicated model.

We have used the asymptotic form $R\sim |t|^{-1/3}$ in the functional
(\ref{e23}). This is justified only as long as the leading asymptotic
expressions $\varrho_{\rm s}\sim |t|^{2/3}$ and $\varrho_0\sim |t|$ are
good approximations. This is the case for
\begin{equation}
\label{e28}
|t|\, \lesssim \, 0.01 \quad (\mbox{range of applicability})\,.
\end{equation}

The fit\cite{fl99} of Eq.\ (\ref{e5}) to the experimental superfluid
provides the temperature dependence of the ratio $R(t)$ in a wider
range. Using this more general function $R(t)$, our approach might be
extended into the less asymptotic region, for example to $0.1 >
|t|>0.01$, where a GL like approach is still meaningful. The emphasis
of the present work is, however, on the temperature range (\ref{e28}).

\subsubsection{Coordinate dependence of the order parameter}
\label{s3.2.3}

We discuss the coordinate dependence of the order parameter field and
its relation to the condensate and superfluid density.  The order
parameter field may be written as
\begin{equation}
\label{e29}
\Psi({\bf r}) = \left| \Psi({\bf r}) \right|  \hspace*{0.2ex}
\exp [\,{\rm i}\hspace*{0.2ex} \Phi({\bf r})] \ .
\end{equation}
We are thus dealing with two real functions, the amplitude $|\Psi({\bf
r}) |$ and the phase $\Phi({\bf r})$, respectively. These functions are
determined by the field equations that follow from the energy
functional (\ref{e23}) and that are supplemented by suitable boundary
conditions.

According to Eqs.\ (\ref{e3}) to (\ref{e5}) all particles contributing
to the superfluid density adopt the phase $\Phi({\bf r})$.
The common phase factor has, however, no implication on the coordinate
dependences of $\varrho_0  ({\bf r})$ and $\varrho_{\rm coh} ({\bf
r})$.

The order parameter field is directly connected to condensate density,
Eq.\ (\ref{e2}). This implies
\begin{equation}
\label{e30}
\varrho_0  ({\bf r}) = m \left\langle |\Psi({\bf r})|^2 \right\rangle
\,.
\end{equation}
It is not a priori clear how to determine the coordinate dependence
of $\varrho_{\rm coh} ({\bf r})$. Eq.\ (\ref{e22}) fixes the
bulk relation
\begin{equation}
\label{e31}
\varrho_{\rm s} (\mbox{bulk}) =  m \hspace*{0.2ex} R(t) 
\left\langle |\Psi(\mbox{bulk})|^2 \right\rangle 
\end{equation}
between the $\Psi$ field and the superfluid density, only. The
coordinate dependence of $\varrho_{\rm s} (\mbox{\bf r})$ versus that
of $|\Psi({\bf r})|^2$ will be discussed in Sec.\ \ref{s6} for the
important case of the behavior at a boundary.

\subsection{Fluctuations of the phase field}
\label{s3.3}

A superfluid flow with the velocity ${\bf v}_{\rm s} = (\hbar/m)
\nabla\Phi$ constitutes a nonequilibrium excitation of the phase
field $\Phi ({\bf r})$ in Eq.\ (\ref{e29}). At finite temperatures
there will be also statistical excitations of the phase field. In the
following we discuss such thermal excitations or fluctuations. This
discussion leads to a qualitative argument for the existence and the
meaning of the coherence limit $k_{\rm coh}$ in Eq.\ (\ref{e4}).

We assume an order parameter field of the form
\begin{equation}
\label{e32}
\Psi ({\bf r}) = \sqrt{\frac{n_0}{V}}\,
\exp [\,{\rm i}\hspace*{0.2ex}\Phi ({\bf r}) ] \ .
\end{equation}
This means that we do not consider spatial variations of $|\Psi({\bf
r})|$ in this subsection.

The average momentum of the fluctuations of the phase field is denoted
by
\begin{equation}
\label{e33}
k_{\rm fluct} = \langle | \nabla \Phi | \rangle
\ .
\end{equation}
This momentum $k_{\rm fluct}$ will be a function of the temperature.

For the single-particle states of the IBG we use real functions
$\varphi_{\bf k}$. We may then consider the possibility of phase
fluctuations for a low-lying state with $n_{\bf k} \gg 1$, i.\,e.\/
phase fields $\Phi_{\bf k}({\bf r})$ that are introduced by the
replacement
\begin{equation}
\label{e34}
\varphi_{\bf k} \longrightarrow \varphi_{\bf k}\exp
[\,{\rm i}\hspace*{0.2ex}\Phi_{\bf k}({\bf r})] \ .
\end{equation}
Let us first assume that these additional phases vanish, $\Phi_{\bf
k}=0$. In this case, the average kinetic energy $\hbar^2 k_{\rm
fluct}^{\,2}/2 m$ of a condensed particle would exceed that of a
noncondensed particle with $k<k_{\rm fluct}$. The energy sequence of
the single-particle states is, however, a prerequisite of the BEC; the
condensate must be formed by the particles with the lowest energy. In
order to preserve the energy sequence of the low-lying states we
require the phase ordering
\begin{equation}
\label{e35}
\Phi_{\bf k}({\bf r}) = \Phi({\bf r}) \quad \mbox{for}\quad
k \le k_{\rm fluct}
\ .
\end{equation}
This argument does not apply to the states with momenta above
$k_{\rm fluct}$.

It is plausible to assume the phase ordering Eq.\ (\ref{e35}) not only
for statistical but also for nonstatistical excitations. This means
that the coherence limit introduced in Eq.\ (\ref{e4}) should be
identified with the average momentum of the fluctuations:
\begin{equation}
\label{e36}
k_{\rm coh} = k_{\rm fluct}
\ .
\end{equation}

In a first approach, the coherence limit $k_{\rm coh}$ in
Eq.\ (\ref{e5}) can be considered as a parameter that has to be
determined from the fit to the experimental superfluid density. In
contrast to this point of view, we present now theoretical arguments
for the leading asymptotic behavior of $k_{\rm coh}$.

The critical exponents of $\varrho_{\rm s}$ and $k_{\rm fluct}$ may be
derived from the following assumptions:
\begin{enumerate} 
\item
The Landau part $G_{\rm L}$ of the thermodynamic potential has its
well-known form (compatible with the IBG).
\item
The fluctuation part $G_{\rm fluct}$ contains the extra factor
$\varrho_{\rm s}/\varrho_0$ due to the assumed comotion of noncondensed
particles (as compared to a standard GL functional).
\item
The fluctuation part scales in the same way as the Landau part. This
scaling invariance is the decisive physical assumption.
\end{enumerate}

We write $G_{\rm GL} = \int\!d^3r\,g_{\rm GL}$ for the thermodynamic
potential. The first two assumptions lead to an integrand of the form
\begin{equation}
\label{e37}
g_{\rm GL} = g_{\rm fluct}  + g_{\rm L} = 
\frac{\hbar^2}{2 \hspace*{0.2ex} m}\,\frac{\varrho_s}{\varrho_0}\,
|\nabla\Psi|^2 + r\,t\,|\Psi|^2 + u\,|\Psi|^4
\ .
\end{equation}
In contrast to Eq.\ (\ref{e23}), we do not introduce an assumption
about the temperature dependence of the ratio $\varrho_s/\varrho_0$.
The present discussion will rather deduce this temperature dependence.

Asymptotically, the Landau part behaves like
\begin{equation}
\label{e38}
\langle g_{\rm L}\rangle  \propto  r\,t\hspace*{0.2ex} \varrho_0 + 
u \hspace*{0.2ex} \varrho_0^{\,2} \sim  t^2
\ .
\end{equation}
The third assumption, the scaling invariance, implies that $\langle
g_{\rm fluct}\rangle$ must have the same leading $t$ dependence.
Because of $\langle|\nabla\Psi|^2\rangle = \varrho_0 \hspace*{0.2ex} 
k_{\rm fluct}^{\,2}$ we have $\langle g_{\rm fluct}\rangle \propto
\varrho_{\rm s} \hspace*{0.2ex} k_{\rm fluct}^{\,2}$. If this term
scales with $t^2$ the critical exponent of $k_{\rm fluct}$ must be
{\em smaller}\/ than 1. This implies that $\varrho_{\rm coh}\propto
k_{\rm coh} = k_{\rm fluct}$ is asymptotically large compared to
$\varrho_{\rm 0} \sim |t|$. Therefore,
\begin{equation}
\label{e39}
\varrho_{\rm s} = \varrho_{\rm 0} + \varrho_{\rm coh} \sim \varrho_{\rm
coh} \sim k_{\rm fluct} \ .
\end{equation}
The last step follows from Eqs.\ (\ref{e11}) and  (\ref{e36}). This
leads to
\begin{equation}
\label{e40}
\langle g_{\rm fluct}\rangle \propto 
\varrho_s \hspace*{0.2ex} k_{\rm fluct}^{\,2} =
(\varrho_0 + \varrho_{\rm coh}) \, k_{\rm fluct}^{\,2}
\sim  k_{\rm fluct}^{\,3}
\ .
\end{equation}
The scaling assumption $\langle g_{\rm fluct}\rangle \sim
\langle g_{\rm L}\rangle \sim  t^2$ implies then
\begin{equation}
\label{e41}
k_{\rm fluct}\sim |t|^{2/3}
\ .
\end{equation}
According to Eq.\ (\ref{e39}) the superfluid density has the same
critical exponent, $\varrho_s\sim k_{\rm fluct}\sim |t|^{2/3}$.

To summarize: The assumption of the coherent comotion (leading to
$\varrho_{\rm coh} \propto k_{\rm coh} = k_{\rm fluct}$) and the scaling
of the fluctuation part ($\langle g_{\rm fluct}\rangle \propto
\varrho_{\rm coh}\,k_{\rm fluct}^{\,2}\propto k_{\rm fluct}^{\,3}$)
with the Landau part ($\langle g_{\rm L}\rangle \sim t^2$) determines
the critical exponent $\nu =1/3$.

\section{Correlation length}
\label{s4}

\subsection{Derivation}
\label{s4.1}

We derive the correlation length in our effective GL model. For this
purpose we add a standard coupling term $-h\cdot \Psi$ to the integrand
of the energy functional (\ref{e23}) where $h$ is a fictitious
external field. The variation of the energy functional with respect to
the $\Psi$ field leads to the field equation
\begin{equation}
\label{e42}
 \frac{a_0\hspace*{0.2ex}\lambda_{\rm c}^{\,2}}{2\pi f\hspace*{0.2ex} 
|t|^{1/3}}\, \Delta\Psi  - 2\,\frac{\Delta c_P}{k_{\rm B}}
\left( t\, \frac{\Psi}{f}  +  \frac{v\, |\Psi|^2
\hspace*{0.2ex} \Psi}{f^2}\right) + h({\bf r}) = 0\ .
\end{equation}
A small external field $h({\bf r}) = \delta h_{\bf k}\exp({\rm i}{\bf
k}\cdot{\bf r})$ will change the field from its equilibrium value
$\langle \Psi \rangle$ to $\Psi = \langle \Psi \rangle + \delta
\Psi_{\bf k}\exp({\rm i}{\bf k}\cdot{\bf r})$. We insert these
expressions for $h({\bf r})$ and  $\Psi({\bf r})$ into the field
equation. In zeroth order in $\delta\Psi$ this yields the equilibrium
value $\langle |\Psi|^2 \rangle$. In first order in $\delta\Psi$ we
obtain
\begin{equation}
\label{e43}
\left[  \frac{a_0\hspace*{0.2ex}\lambda_{\rm c}^{\,2}}
{2\pi f \hspace*{0.2ex} |t|^{1/3}} \, k^2  +  
2\,\frac{\Delta c_P}{k_{\rm B}} \left( \frac{t}{f}  + 
3\, \frac{v\, \langle |\Psi|^2\rangle }{f^2} \right)\right] 
\delta\Psi_{\bf k} =  \delta h_{\bf k} \ .
\end{equation}
This susceptibility of the system with respect to an external field is
given by
\begin{equation}
\label{e44}
\chi (k) = \frac{\delta \Psi_{\bf k}}{ \delta h_{\bf k}} =
 \frac{ \chi (0)}{ 1+ k^2\,\xi^2} \ .
\end{equation}
The $k$ dependence displayed in the last expression follows from
Eq.\ (\ref{e43}). It implies
\begin{equation}
\label{e45}
h({\bf r}) = h_0 \,\delta ({\bf r}) \quad \longrightarrow \quad
\delta \Psi ({\bf r}) \propto \exp(-r/\xi)\ .
\end{equation}
This means that the quantity $\xi$ introduced in Eq.\ (\ref{e44})
defines the {\em correlation length}\/. Comparing Eqs.\ (\ref{e43})
and (\ref{e44}) yields
\begin{equation}
\label{e46}
\xi^2 =   \frac{
\frac{a_0\hspace*{0.2ex}\lambda_{\rm c}^{\,2}}{2\pi 
\hspace*{0.2ex} |t|^{1/3}} } { 2\,\frac{\Delta c_P}{k_{\rm B}}
\left( t + 3\, \frac{v\hspace*{0.2ex} 
\langle |\Psi|^2\rangle }{f}\right)}\ .
\end{equation}
Inserting the equilibrium values $\langle |\Psi|^2\rangle = 0$ for
$t<0$ and $\langle |\Psi|^2\rangle = f \hspace*{0.2ex} |t|/v$  for
$t<0$ we obtain
\begin{equation}
\label{e47}
\xi(t) = \left\{\begin{array}{lll}
\xi^+ (t) = \xi_0 \,|t|^{-2/3} && (t>0)\\[0mm]
\xi^- (t) = \xi_0 \,|t|^{-2/3}/\sqrt{2}&& (t<0)
\end{array}\right. 
\end{equation}
where 
\begin{equation}
\label{e48}
\xi_0 = \frac{\lambda_{\rm c}}{\sqrt{4\pi}}
\,\sqrt{\frac{a_0}{\Delta c_P/k_{\rm B}}} 
\approx 1.53\,\mbox{\AA}\ .
\end{equation}
For the numerical value we used Eqs.\ (\ref{e24}) and (\ref{e25}). The
result is independent of the parameter $f$.

The $\Psi$ theory by GS leads to Eqs.\ (\ref{e47}) and (\ref{e48}),
too, but with an additional factor $[(3+M)/3]^{1/2}$ for $\xi^+$ and
$[(3+M)/(3+3M)]^{1/2}$ for $\xi^-$; here $M$ is an extra parameter.
For $M=0$ the results by GS and the present model coincide. For the
favored $M$ value by GS ($M=0.6\pm 0.3$ according to Eq.\ (\ref{e20}) in
Ref.\onlinecite{gs88}) the GS values differ from ours by about 10\%{}.

The validity of the result  Eq.\ (\ref{e47}) is restricted to the
temperature range $|t|\lesssim 0.01$, Eq.\ (\ref{e28}). Outside this
asymptotic region we expect the following behavior: For $|t|> 0.2$ we
find (Fig.\ \ref{fig1}) $\varrho_{\rm s}\approx \varrho_0$
corresponding to $R\approx 1$ in Eq.\ (\ref{e22}). This implies that in
the temperature range from $|t|=0.01$ to $|t| = 0.2$ the correlation
length will change its behavior from $\xi^-\sim |t|^{-2/3}$ to
$\xi^-\sim |t|^{-1/2}$.

\subsection{Critical fluctuations}
\label{s4.2}

A standard GL ansatz (in $d=3$ dimensions) breaks down for $|t|\to 0$
because the ratio of the fluctuating to the equilibrium field diverges.
In this section we will show that this problem does not exist in our
effective GL model. This is due to the scaling invariance (Sec.\
\ref{s3.3}) of our energy functional. The following discussion proceeds
along the lines followed in a standard GL approach (see, for example,
Ref.\onlinecite{ll}).

Equation (\ref{e43}) determines the deviation $\delta\Psi$ from the
equilibrium due to an external field. Deviations from equilibrium
occur also due to fluctuations. The fluctuations of the
thermodynamic potential are of the size $\delta G =  {\cal O}(k_{\rm
B}T)$. This determines the size of the thermal fluctuations
$\delta \Psi_{\rm therm}$ of the order parameter field.

We write $\Psi =\langle \Psi\rangle + \delta \Psi $ and  expand $G_{\rm
GL} = \langle G_{\rm GL}\rangle + \delta G_{\rm GL} $ into powers of
$\delta \Psi$. The term linear in $\delta \Psi$ vanishes because we
expand around the equilibrium value. Therefore, $\delta G_{\rm GL}$
starts with the quadratic term:
\begin{equation}
\label{e49}
\frac{\delta G_{\rm GL}}{k_{\rm B}T_\lambda} =
 \int \!d^3r \,\frac{1}{2}\frac{\partial^2 g}{\partial
|\Psi|^2} \, \left| \delta \Psi\right|^2 \ .
\end{equation}
Here $g$ denotes the integrand in Eq.\ (\ref{e23}). The derivative has
to be taken at the equilibrium value:
\begin{eqnarray}
\label{e50}
\frac{\partial^2 g}{\partial |\Psi|^2} &=&
 2\,\frac{\Delta c_P}{k_{\rm B}}
\left( \frac{t}{f} + 
\frac{3 \hspace*{0.2ex} v
\hspace*{0.2ex} \langle |\Psi|^2\rangle }{f^2}\right)
\nonumber \\[1.5mm] &=&
\frac{2}{f} \frac{\Delta c_P}{k_{\rm B}} \,|t| \cdot
\left\{ \begin{array}{ccc}
1 && (t>0) \\[0mm] 2 && (t<0)
\end{array}\right. \ .
\end{eqnarray}
The typical range of a thermal fluctuation $\delta \Psi_{\rm therm}$
is given by the correlation length $\xi(t)$. For the following estimate
we may, therefore, replace the $\int\!d^3r$ in Eq.\ (\ref{e49}) by
$\xi^3$. Using this approximation and $\delta G_{\rm GL,\,therm} =
{\cal O}(k_{\rm B}T_\lambda)$, Eq.\ (\ref{e49}) becomes
\begin{equation}
\label{e51}
\frac{\delta G_{\rm GL}}{k_{\rm B}T_\lambda} \approx \frac{\xi^3}{2}
\frac{\partial^2 g}{\partial |\Psi|^2} \, 
\left| \delta \Psi_{\rm therm} \right|^2 = {\cal O}(1)\ .
\end{equation}
Inserting Eq.\ (\ref{e50}) and omitting factors of the order 1 we obtain
\begin{equation}
\label{e52}
\left| \delta  \Psi_{\rm therm}\right|^2 = 
\frac{{\cal O}(1)}{|t|\hspace*{0.2ex} \xi(t)^3 } \sim |t| \,.
\end{equation}
For the last step we used Eq.\ (\ref{e47}), $\xi \sim |t|^{-2/3}$. 
The applicability of a GL approach is restricted to temperatures where
the fluctuations are small (or at least not large) compared to the mean
value of the field. The relevant ratio is
\begin{equation}
\label{e53}
\frac{| \delta  \Psi_{\rm therm}|^2 }
{ \langle | \Psi|^2\rangle } \sim
\left\{ \begin{array}{lll}
\mbox{const.} && \mbox{present model} \\[0.8mm]
|t|^{-1/2} && \mbox{GL}
\end{array}\right. \,.
\end{equation}
In a standard GL approach the ratio diverges for $|t|\to 0$; this
excludes the application of the model in the asymptotic region. In our
effective GL model the ratio remains finite for $|t|\to 0$. This is
the same behavior as in a standard GL approach in $d=4$ dimensions.

\section{Transition temperature in helium film}
\label{s5}

We consider a helium film on the $y$-$z$-plane extending from $x=0$ to
$x=D$. We look for a nonzero solution $\Psi(x)$ of the field equation
that obeys the boundary conditions
\begin{equation}
\label{e54}
\Psi(0) = \Psi (D) = 0 \ .
\end{equation}
This implies that a nonvanishing $\Psi(x)$ is inhomogeneous. Due to
the kinetic energy term in Eq.\ (\ref{e23}) such a nonvanishing
$\Psi(x)$ becomes possible only at a temperature $T_{\lambda}(D)$ below
$T_{\lambda} = T_{\lambda}(\infty)$. We calculate this transition
temperature $T_{\lambda}(D)$ as a function of the film thickness $D$.

In an experiment the plane $x=0$ might be a solid wall, and the
plane $x=D$ may define the free surface of the helium film. In this
situation one has a solid layer (thickness $d$) of helium at the wall
so that the first boundary condition actually reads $\Psi(d)=0$. This
point amounts to replacing $D$ by $D-d$ and is ignored in the
following. For a discussion of the appropriate boundary condition
at the free surface we refer to GS\cite{gs82}.

We consider the field equation Eq.\ (\ref{e42}) for $h({\bf r})=0$ and
$t<0$. We introduce the quantity $\xi_0$ of Eq.\ (\ref{e48}) and
specialize to the one dimension of interest:
\begin{equation}
\label{e55}
\xi_0^{\,2}\hspace*{0.2ex} |t|^{-4/3}\, \frac{d^2\Psi(x)}{dx^2} + 
\Psi  \left(1- \frac{v\,|\Psi|^2}{f\hspace*{0.2ex}  |t|}\right) = 0  \ .
\end{equation}
For determining the {\em onset}\/ of a nonvanishing $\Psi$, the
$\Psi^3$ term may be neglected. The solution of the remaining
differential equation is $\Psi = A\hspace*{0.2ex}
\sin(x\hspace*{0.2ex} |t|^{2/3}/\xi_0 +\alpha)$ where $|A|^2 \ll
f|t|/v$. The boundary condition $\Psi(0) = 0$ implies $\alpha =0$,
and $\Psi (D)=0$ leads to $D\hspace*{0.2ex} |t|^{2/3}/\xi_0 =
n\hspace*{0.2ex} \pi$ where $n$ is a positive integer. We are looking
for the highest temperature $T$ or the smallest $|t|$ value ($t<0$) for
which such a solution exists. That means we have to take the $n=1$
solution for which the boundary condition yields
\begin{equation}
\label{e56}
\pi\hspace*{0.2ex} \xi_0\hspace*{0.2ex} |t|^{-2/3} = D \,.
\end{equation}
This defines the maximum $T$ value for a nontrivial solution. For a
more elaborate discussion of the underlying mathematics we refer to
Ref.\onlinecite{gp58}.  Using $t(D) = [T_{\lambda}(D) -
T_{\lambda}(\infty)] / T_{\lambda}(\infty)$ we resolve Eq.\ (\ref{e56})
for $T_{\lambda}(D)$:
\begin{eqnarray}
\label{e57}
T_{\lambda}(D) &=&  T_{\lambda}(\infty) \left[ \, 1 - \left(
\frac{\pi \hspace*{0.2ex} \xi_0}{D} \right)^{\! 3/2} \, \right]
\nonumber\\[1mm]
&\approx &
T_{\lambda}(\infty) \left[ \, 1 - \left(
\frac{4.81\,\mbox{\AA}}{D} \right)^{\! 3/2} \, \right]\ .
\end{eqnarray}
Here $T_{\lambda}(\infty)$ denotes the transition temperature of the
bulk system. The numerical value follows from Eq.\ (\ref{e48}).

As discussed by GS, the experimental and theoretical situation in a
helium film is more delicate. According to GS\cite{gs88} there is a
caloric transition, and at a slightly lower temperature another
transition to superfluidity. The condition for the second transition is
somewhat different from that one discussed in this section. As shown by
Kosterlitz and Thouless\cite{kt73}, superfluidity sets in when the
superfluid density (averaged over the film thickness) exceeds a certain
minimum value. Evaluating this condition in our model leads to an
expression of the form (\ref{e57}), too, but with a slightly different
length parameter (instead of $\pi\hspace*{0.2ex} \xi_0$).

For the comparison with the experimental findings we refer again to
GS\cite{gs88}. One finds indeed transition temperature shifts that are
proportional to $D^{-3/2}$, and length scales similar to the
calculated one.

\section{Density profiles at a boundary}
\label{s6}

\subsection{Introduction}
\label{s6.1}

We consider a helium film with a thickness $D$ much larger than the
correlation length $\xi$. At the wall the superfluid density will raise
from zero to its bulk value $\varrho_{\rm s}(\mbox{bulk})$ (for
simplicity we ignore again the solid layer at the wall). Similarly, the
superfluid density will fall off again towards the free surface. 

The following discussion is restricted to the density profile at one
boundary. This means that we determine how the density profile
\begin{equation}
\label{e58}
f(x) = \frac{\varrho_{\rm s}(x)}{ \varrho_{\rm s}(\mbox{bulk})} 
\end{equation}
raises from zero at $x=0$ to its bulk value. This corresponds to the
boundary conditions
\begin{equation}
\label{e59}
f(0) = 0 \quad \mbox{and}\quad
f(\infty )=1 \ .
\end{equation}
Third sound measurements determine the value $\overline{\varrho_{\rm
s}}$ of the superfluid density averaged over the film thickness $D$.
By $\overline{\varrho_{\rm s}} = \varrho_{\rm s}(\mbox{bulk}) (D -
2\hspace*{0.2ex} \xi_{\rm heal})/D$ one defines the {\em healing
length}\/ $\xi_{\rm heal}(t)$ (for a more comprehensive discussion see
\S\,41 in Ref.\onlinecite{pu74}). For a given profile $f(x)$ this
healing length may be calculated by
\begin{equation}
\label{e60}
\xi_{\rm heal} = \lim_{a\to \infty} 
\left( a - \int_0^a \! dx\,f(x)\right) \ .
\end{equation}

We approach the question of the density profile in two different ways:
\begin{enumerate}
\item 
In the considered asymptotic regime ($|t|\lesssim 0.01$) the density
$\varrho_{\rm coh}$ is the predominant contribution to the superfluid
density. The particles forming $\varrho_{\rm coh}$ have a distribution
of momenta up to about $k_{\rm coh}$. This leads to a healing length
$\xi_{\rm heal,1}
\sim 1/k_{\rm coh}\sim |t|^{-2/3}$.
\item 
We solve the field equation (\ref{e55}) with the boundary conditions
Eq.\ (\ref{e59}).  This yields the profile of condensate density
$\varrho_0$, Eq.\ (\ref{e30}), and to a healing length $\xi_{\rm
heal,2} \sim \xi_0 \hspace*{0.2ex} |t|^{-2/3}$.
\end{enumerate}
The composition (\ref{e5}) of the superfluid density leads thus to the
prediction that the boundary profiles of the densities $\varrho_{\rm
s}$ and $\varrho_0$ are different.

\subsection{Profile of the superfluid density}
\label{s6.2}

For $|t|\ll 1$ the superfluid density may be written as
\begin{equation}
\label{e61}
\varrho_{\rm s} \approx \varrho_{\rm coh} \propto 
\int_0^{k_{\rm coh}} \! dk\,k^2\,
\langle n_k\rangle \,|\varphi_{\bf k}({\bf r})|^2
\ .
\end{equation}
The $\varphi_{\bf k}({\bf r})$ are the real single particle functions
of lowlying noncondensed states. For $|t|\ll 1$ the occupation numbers
(\ref{e6}) are $\langle n_k\rangle \approx 1/(\kappa ^2 + \tau^2)$.
Because of $\tau^2\ll \kappa_{\rm coh}^{\,2}$ we may write
$\kappa^2/(\kappa ^2 + \tau^2) \approx 1$. At the boundary plane $x=0$
all single particle functions must vanish, i.\,e.\/ $|\varphi_{\bf
k}({\bf r})|^2 \propto \sin ^2 (k_x\hspace*{0.2ex} x)$. Since $k^2 =
k_x^{\,2} + k_y^{\,2}+ k_z^{\,2}$ is limited by $k_{\rm coh}^{\,2}$, we
assume that $k_x^{\,2}$ is limited by $k_{\rm coh}^{\,2}/3$. Putting
these (partly approximate) steps together yields
\begin{eqnarray}
\label{e62}
\varrho_{\rm s}(x) &\propto& \int_0^{k_{\rm coh}/\sqrt{3}}
\!
dk_x\, \sin^2(k_x \hspace*{0.2ex} x) 
\nonumber\\[1mm] 
&=& \frac{k_{\rm koh}}{2\,\sqrt{3}} -
 \frac{\sin\left(2 \hspace*{0.2ex} k_{\rm coh}\hspace*{0.2ex} 
 x/\sqrt{3}\right)}{4\hspace*{0.2ex} x}
\ .
\end{eqnarray}
Including all constants, the first term on the r.\,h.\,s.\ equals the
bulk value; it is obtained as the limit of Eq.\ (\ref{e62}) for
large $x$. Therefore, we may read off the density profile:
\begin{equation}
\label{e63}
\frac{\varrho_{\rm s}(x)}{ \varrho_{\rm s}(\mbox{bulk})} = 1 - 
\frac{\sin\left(2 \hspace*{0.2ex} k_{\rm coh}\hspace*{0.2ex} 
x/\sqrt{3}\hspace*{0.2ex}\right)}
{ 2 \hspace*{0.2ex} k_{\rm coh}\hspace*{0.2ex}  x/\sqrt{3}} = f_1(x)
\ .
\end{equation}
We evaluate the healing length (\ref{e60}) for this profile,
\begin{equation}
\label{e64}
\xi_{\rm heal,\,1} =
\frac{\pi \,\sqrt{3}}{4 \hspace*{0.2ex} k_{\rm coh}}
= \frac{\sqrt{3}}{2\,\zeta(3/2)}
\frac{\lambda_{\rm c}}{a_0} \,|t|^{-2/3} 
= 0.82\,\mbox{\AA}\,|t|^{-2/3} 
\ .
\end{equation}
We used Eqs.\ (\ref{e7}), (\ref{e13}) and (\ref{e14}) for the second
step, and Eqs.\ (\ref{e24}) and (\ref{e25}) for the numerical value.

\subsection{Profile of the condensate density}
\label{s6.3}
Using the dimensionless quantities $\psi = \Psi(x)/\Psi(\mbox{bulk})$
with $\Psi(\mbox{bulk})=\sqrt{ f|t|/v\,}$ and $y = (x/\xi_0)\,
|t|^{2/3}$, the field equation (\ref{e55}) reads
\begin{equation}
\label{e65}
\psi''(y) + \psi(y) \left[1 -\psi(y)^2\right] = 0
\ .
\end{equation}
The boundary conditions Eq.\ (\ref{e59}) become $\psi(0)=0$ and
$\psi(\infty)=1$.  The solution of Eq.\ (\ref{e65}) obeying these
boundary conditions is $\psi(y) = \tanh (y/\sqrt{2})$. Returning to the
original variables  we obtain the profile of the condensate density
(\ref{e30}),
\begin{equation}
\label{e66}
\frac{\varrho_0(x)}{ \varrho_0(\mbox{bulk})} = 
\tanh^2\left( \frac{x}{\sqrt{2}\,\xi_0\hspace*{0.2ex} |t|^{-2/3}
}\right)  =  f_2(x)
\ .
\end{equation}
Inserting this profile into Eq.\ (\ref{e60}) yields the healing length 
\begin{equation}
\label{e67}
\xi_{\rm heal,\,2} =
\sqrt{2}\,\xi_0 \hspace*{0.2ex} |t|^{-2/3}
= 2.16 \,\mbox{\AA}\,|t|^{-2/3} 
\ .
\end{equation}
This differs by a factor of 2 from the correlation length $\xi^-(t)$ .

\subsection{Discussion}
\label{s6.4}

\begin{figure}[h]
\begin{center}
\epsfxsize=8cm
\epsfbox{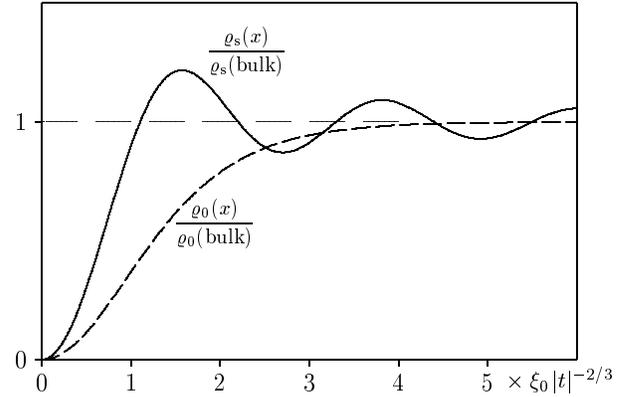}
\end{center}
\caption[]{\label{fig2}Superfluid density $\varrho_{\rm s}(x)$ and the
condensate density $\varrho_0(x)$ as functions of the distance $x$
from a boundary. The distance $x$ is measured in units of
$\xi_0\hspace*{0.2ex} |t|^{-2/3}$. For large $x$ the densities approach
their bulk values.}

\end{figure}

The calculated density profiles of the superfluid and condensate
density, Eqs.\ (\ref{e59}) and (\ref{e66}), respectively, are depicted
in Fig.\ \ref{fig2}. These results refer ---as most results in this
paper--- to the asymptotic region $|t|\lesssim 0.01$, Eq.\ (\ref{e28}).

The occurrence of different profiles does not alter the calculation
of Sec.\ \ref{s5}: For a superfluid flow we must have a nonvanishing
order parameter field. The condition obtained in Sec.\ \ref{s5} is
sufficient to accommodate both, $\varrho_0$ and $\varrho_{\rm coh}$,
within the film.

The experimental result for the healing length cited in Eq.\ (41-47) of
Ref.\onlinecite{pu74} is $\xi^{\rm exp}_{\rm heal} \approx
1.55\,\mbox{\AA} \,|t|^{-2/3}$. The question is which one of the
calculated results, (\ref{e64}) or (\ref{e67}), should be compared with
the experimental one. Both results scale correctly.  Relative to the
experimental prefactor, one theoretical prefactor is about 50\% to
large, in the other case about 50\% too small. Because of $\varrho_{\rm
coh}(\mbox{bulk})\gg \varrho_0(\mbox{bulk})$ (for $|t|\lesssim 0.01$)
one might be inclined to consider the correlation length (\ref{e64}) as
the relevant quantity. However, (i) the macroscopic $\Psi$ field defines
the macroscopic phase field necessary for superfluidity, and (ii) the
simple averaging procedure (\ref{e60}) might not be appropriate for a
profile like $f_1(x)$. In view of this more involved situation, we
restrict ourselves to the statement that the theoretical results are in
a reasonable vicinity of the experimental one.

A direct experimental verification of the predicted profiles shown in 
Fig.\ \ref{fig2} appears hardly feasible. The $\varrho_{\rm s}$ profile
may have some influence on third sound calculations the results of
which might then be tested. The condensate density is already in the
bulk system\cite{sn92} a hard to access quantity.

The different profiles calculated here are obviously in variance with
the GS approach. It is a future task to find differing predictions of
our model that might be readily subject to an experimental test.

\section{Concluding remarks}
\label{s7}

We have proposed an effective Ginzburg-Landau model that is compatible
with the bulk properties of liquid helium. We have investigated the
basic properties of this model and its most simple applications for
finite geometries.

Our model is based on the IBG and the assumption of a coherent comotion
of condensed and noncondensed particles. The occurrence of fractional
powers in the energy functional can be made plausible on this basis.
The foundation of our model is thus different from the $\Psi$ theory by
GS\cite{gs76,gs82,gs88}.

The results obtained so far are mostly similar to that of the
$\Psi$ theory by GS\cite{gs76,gs82,gs88}. This is to be expected for
other applications, too. The results for the density profiles at a
boundary are, however, novel and in variance with the $\Psi$ theory.

There is a variety of further problems that might be treated in our
approach (and that have already been treated in the $\Psi$ theory),
like the influence of external fields or of impurities (ions) and the
description of superfluid flow in various geometries.

The $\Psi$ theory has been generalized\cite{gs88} in order to
incorporate (i) time-dependent problems in the field equation, (ii)
dissipation effects and (iii) problems with a nonvanishing normal
velocity. These generalizations correspond to additional terms in the
functional of the thermodynamic potential. All these generalization can
be introduced in our model in a straight-forward way, too.



\begin{thebibliography}{99}
\bibitem[*]{email} Electronic address: fliessbach@physik.uni-siegen.de
\bibitem{gp58} V. L. Ginzburg and L. P. Pitaevski\u{\i}, Zh. Eksp.
Teor. Fiz. {\bf 34}, 1240 (1958) [Sov. Phys. JETP {\bf 7}, 858 (1958)]
\bibitem{gs76} V. L. Ginzburg and A. A. Sobyanin, Usp. Fiz. Nauk {\bf
120}, 153 (1976) [Sov. Phys. Usp. {\bf 19}, 773 (1976)]
\bibitem{gs82} V. L. Ginzburg and A. A. Sobyanin, J. Low Temp. Phys. 
         {\bf 49}, 507 (1982)
\bibitem{gs88} V. L. Ginzburg and A. A. Sobyanin, Usp. Fiz. Nauk {\bf
154}, 545 (1988) [Sov. Phys. Usp. {\bf 31}, 289 (1988)]
\bibitem{fe53} R. P. Feynman, Phys. Rev. {\bf 91}, 1291 (1953).
\bibitem{lo54} F. London, {\em Superfluids} (Wiley, New York
        1954), Vol. II.
\bibitem{pu74} S. J. Putterman, {\em Superfluid Hydrodynamics} (North
        Holland, London 1974).
\bibitem{fl99} T. Fliessbach, Phys. Rev. B\,{\bf 59}, 4334 (1999) or
cond-mat/9901175
\bibitem{no90} P. Nozi\`ere, {\em The Theory of Quantum Liquids, Volume
      II: Superfluid Bose Liquids} (Addison-Wesley, Redwood City, 1990).
\bibitem{sn92} W. M. Snow, Y. Wang, and P. E. Sokol, Europhys. Lett.
         {\bf 19}, 403 (1992).
\bibitem{ah71} G. Ahlers, Phys. Rev. {\bf A\,3}, 696 (1971)
\bibitem{ll} L. D . Landau and E. M. Lifschitz, {\em Statistische 
            Physik}\/, 7. ed., Akademie-Verlag Berlin 1987, \S{}\,146
\bibitem{kt73}J. M. Kosterlitz and D. J. Thouless, J. Phys. {\bf C6},
              1181, (1973)
\end{thebibliography}
\end{document}